\documentclass[useamsfonts]{pasj00}
\usepackage{mathrsfs}
\usepackage{graphicx,epsfig} 
\begin{document}
\hyphenation {Schwarz-schild}
\hyphenation {Abra-mo-wicz}

\SetRunningHead{M.~A. Abramowicz et al.}{Non-linear resonance in LMXBs}
\Received{2002/12/22}
\Accepted{2003/01/15}

\title{Non-linear resonance in nearly geodesic motion\\
 in low-mass X-ray binaries}

\author{M.\,A.~\textsc{Abramowicz}\altaffilmark{1,2},
V.~\textsc{Karas}\altaffilmark{3,$\ddag$},
W.~\textsc{Klu{\'z}niak}\altaffilmark{2,4,5,$\S$},
W.\,H.~\textsc{Lee}\altaffilmark{2,6}, and
P.~\textsc{Rebusco}\altaffilmark{2,7}}

\affil{$^1$~Astrophysics Department, Chalmers University,
 S-412\,96 G{\"o}teborg, Sweden}
\affil{$^2$~Scuola Internazionale Superiore di Studi Avanzati (SISSA),
 via Beirut 2-4, I-34\,014 Trieste, Italy}
\affil{$^3$~Astronomical Institute, Charles University Prague,
 V~Hole{\v s}ovi{\v c}k{\'a}ch 2, CZ-180\,00 Praha, Czech Republic}
\affil{$^4$~Institute of Astronomy, Zielona G{\'o}ra University, Lubuska 2,
 P-65\,265 Zielona G{\'o}ra, Poland}
\affil{$^5$~CESR, 9, ave.\ Colonel-Roche, F-31028 Toulouse Cedex 4, France}
\affil{$^6$~Instituto de Astronom{\'{\i}}a,
 UNAM, Apdo.\ Postal 70-264 Cd.~Universitaria, DF\,04510, Mexico}
\affil{$^7$~Department of Physics, Trieste University, I-34\,127 Trieste, Italy}
\maketitle
\begin{abstract}
We have explored the ideas that parametric resonance affects nearly
geodesic motion around a black hole or a neutron star,
and that it may be relevant to the high frequency (twin)
quasi-periodic oscillations occurring in some low-mass X-ray binaries. 
We have assumed the particles or fluid elements of
an accretion disc to be subject to an isotropic perturbation of
a hypothetical but rather general form. We find that the
parametric resonance is indeed excited close to the radius where
epicyclic frequencies of radial and meridional oscillations are in a
$2:3$ ratio. The location and frequencies of the
highest amplitude excitation vary with the strength of the perturbation.
These results agree with actual frequency ratios of twin kHz QPOs that have 
been reported in some black hole candidates, and they may be consistent also 
with correlation of the twin peaks in Sco X-1.
\end{abstract}

\renewcommand{\thefootnote}{\fnsymbol{footnote}}
\setcounter{footnote}{3}
\footnotetext{On leave of absence at the Center for Particle Physics,
Institute of Physics of the Czech Academy of Sciences.}
\setcounter{footnote}{4}
\footnotetext{On leave of absence from the Copernicus Astronomical Center 
of the Polish Academy of Sciences.}
\renewcommand{\thefootnote}{\arabic{footnote}\,}
\setcounter{footnote}{0}

\section{Introduction}
Pairs of high frequencies, known as kilohertz quasi-periodic oscillations
(kHz QPOs), have been detected
in the X-ray emission of low-mass X-ray binaries (LMXBs)
for more than 20 neutron stars and several black holes.
It has been suggested that a non-linear resonance within an accretion 
disc in a
general-relativistic space-time metric plays a role in the excitation of
the two oscillations (Abramowicz \& Klu\'zniak 2001, Klu\'zniak \& 
Abramowicz 2001).\footnote{Non-linear resonance 
has been favored as an explanation of QPOs also by Titarchuk (2002)
in a rather different (``transition layer'') model. In another context,
a tidally forced, Newtonian, non-linear vertical resonance in 
accretion discs has been discussed
for close binary systems (Lubow 1981; Goodman 1993).}

It is now recognized that the frequencies of the two
peaks in the power spectrum of X-ray variability in black holes are in
rational ratios, $\omega_{\rm 1}/\omega_{\rm 2} = m:n$, with $m:n=2:3$
for two sources, and $m:n=3:5$ in a third source (Abramowicz \&
Klu{\'z}niak 2001; Klu{\'z}niak \& Abramowicz 2002;  Remillard
et al.\ 2002, Abramowicz et al.\ 2002a).
These black hole QPOs are thought to have stable 
frequencies, and until the discovery of
the puzzling rational ratios they have been thought to correspond to a
trapped g-mode or c-mode of disc oscillation in the Kerr metric 
(Okazaki et al.\ 1987, Wagoner 1999, Wagoner et al.\ 2001,
Kato 2001).

In neutron star sources, the twin kHz QPO frequencies are known to
vary considerably and to be mutually correlated. In the context of 
disc oscillations, two frequencies varying in this fashion can
be explained as the fundamental and the first harmonic 
of the non-axisymmetric ({\sl{}m}=1) g-mode (Kato 2002, 2003).
In Sco X-1, a prototypical {\sf{}Z}-source, the slope
of the correlation line is somewhat steeper than 2/3, but it is not
uncommon to find the source in a state when $\omega_{\rm 1}\approx600$~Hz,
and $\omega_{\rm 2}\approx900$~Hz, i.e., at about $2:3$ ratio 
(Abramowicz et al.\ 2002b).

It would appear that the $2:3$ frequency ratio is common to 
neutron star and black hole
systems. Klu\'zniak \& Abramowicz (2002), and Abramowicz \& Klu\'zniak
(2003) suggest that the $2:3$ ratio follows from the properties of
parametric resonance between two eigenfrequencies, of which one is always
lower than the other. Specifically, two modes of disc oscillation are
thought possible with frequencies close to the two epicyclic frequencies of
free orbital motion. Within general relativity, the radial epicyclic frequency
is generically lower than the meridional one, their ratio varying from
one to zero as the radius of circular orbits decreases from infinity
down to that of the marginally stable orbit.

The observed rational ratios of frequencies may follow from fundamental
properties of both strong gravity and a non-linear resonance between radial
and vertical oscillations in accretion discs with fluid lines that are
nearly geodesic, nearly circular, and nearly planar. In this paper we
investigate a simple mathematical model of such motions. To this aim, 
we discuss the properties of parametric resonance in the Paczy\'nski-Wiita 
(1980) model of the Schwarzschild metric.

\begin{figure}[!tb]
\begin{center}
\FigureFile(0.48\textwidth,0.48\textwidth){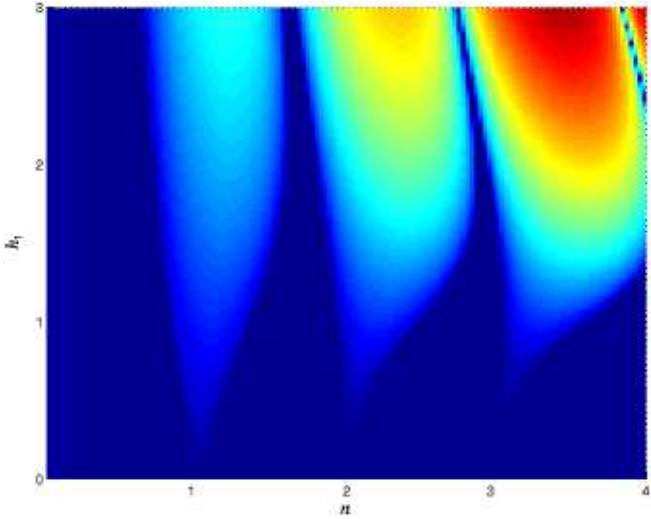}
\end{center}
\caption{Instability regions of the Mathieu equation (\ref{8})
with the amplitude of variation of the eigenfrequency $h_1$,
$n\equiv2\omega_\theta/\omega_r$, and $\lambda=0$.
Levels of shading indicate different values of the growth rate
of $\delta\theta(t)$.
Three tongues of instability, of the order $n=1$, 2, and 3,
are clearly visible. The equations (\ref{2.6a})--(\ref{2.6b})
considered in this paper have similar regions of instability.
Note that the range of frequencies for which the instability 
develops increases with $h_1$.}
\label{fig2}
\end{figure}

\section{Nearly-geodesic motion}
Let us consider a fluid in which the flow lines are only slightly 
non-circular, and slightly off the $\theta = \pi/2$ symmetry plane.
In spherical coordinates,
\begin{equation}
r(t) = r_0 + \delta r(t), ~~\theta (t) = {\textstyle\frac{1}{2}}\pi + 
\delta \theta (t), ~~ \phi (t) = \Omega t.
\label{2.5} 
\end{equation}
With accuracy to the third order in $\delta r \ll r_0$ and
$\delta \theta \ll \pi/2$, equations of fluid motion read:
\[
{\ddot {\delta \theta}} + \omega_{\theta}^2 \left [ 1 
+  {{(\omega_{\theta}^2)^{\prime}}\over {\omega_{\theta}^2 }}\;\delta r
+ {1\over 2}{{(\omega_{\theta}^2)^{\prime \prime}}
\over {\omega_{\theta}^2 }}
\;\delta r^2 \right ]\delta \theta
\]
\begin{equation}
\quad + {2 \over r}  \left ( 1 -{{\delta r}\over r}\right ) 
{\dot {\delta \theta}}\,{\dot {\delta r}} 
+ {1\over {6r^2}}\left ( {{\partial^4 {\cal U}}
\over {\partial ~\theta^4}}\right )_{\!\ell}
\delta \theta^3 = f_{\theta},
\label{2.6a} 
\end{equation}

\[
{\ddot {\delta r}} + \omega_r^2 \,\delta r
+ {\textstyle\frac{1}{2}} (\omega_r^2)^{\prime} \,\delta r^2
+ {\textstyle\frac{1}{6}} (\omega_r^2)^{\prime \prime}\,\delta r^3 
\]
\begin{equation}
\quad - r {\left(\dot{\delta \theta}\right)^2} +
 \delta r {\left(\dot {\delta \theta}\right)^2}
 = f_r,
\label{2.6b} 
\end{equation}
							
\begin{equation}
{\dot \ell} = r^2 \sin^2 \theta\; f_{\phi}.
\label{2.6c}
\end{equation}
Here the time derivative is denoted by a dot, the radial
derivative by a prime, and $f_i$ are components of 
a small force of non-gravitational origin
(pressure, viscous, magnetic, or other). The epicyclic eigenfrequencies
$\omega_{\theta}$ and $\omega_r$ are defined, in terms of the effective
potential ${\cal U} = \Phi (r, \theta) + {\ell}^2/(2 r^2 \sin^2 \theta)$
and the specific angular momentum ${\ell} = {\dot \phi} r^2 \sin^2 \theta$,
by

\begin{equation}
\omega_{\theta}^2 = 
\frac{1}{r^2}
\left({{\partial^2 {\cal U}}\over {\partial ~\theta^2}}\right )_{\!\ell},
\quad
\omega_r^2 = 
\left({{\partial^2 {\cal U}}\over {\partial ~r^2}}\right )_{\!\ell}.
\label{5}
\end{equation}
Derivation of these equations assumes equatorial plane symmetry
for the gravitational potential $\Phi$. Here we adopt a spherically
symmetric potential\footnote{See Paczy{\'n}ski \&
Wiita (1980). This form is known to be a convenient model for the
external gravitational field of a non-rotating black hole or a neutron star;
it captures the essential feature of Einstein's strong gravity relevant
here, i.e., the fact that $\omega_r(r) < \omega_{\theta}(r)$, and 
$\omega_r = 0$ at the marginally stable orbit.}
\begin{equation}
\Phi(r) = - {{GM} \over {r - r_{_{\rm G}}}}, \quad
r_{_{\rm G}} ={{2GM}\over c^2}.
\label{6}
\end{equation}
From eqs.~(\ref{5}) and (\ref{6}) it follows

\begin{equation} 
\omega_r^2 = 
{{GM} \over {\left(r - r_{_{\rm G}}\right)^{3}}}
 \left(1-\frac{r_{\rm ms}}{r}\right)
< \omega_\theta^2 = \frac{GM}{r (r-r_{_{\rm G}})^2},
\end{equation}
where $r_{\rm ms} = 3 r_{_{\rm G}}$ is the marginally stable orbit. 

\begin{figure*}[!tb]
\begin{center}
\FigureFile(0.48\textwidth,48\textwidth){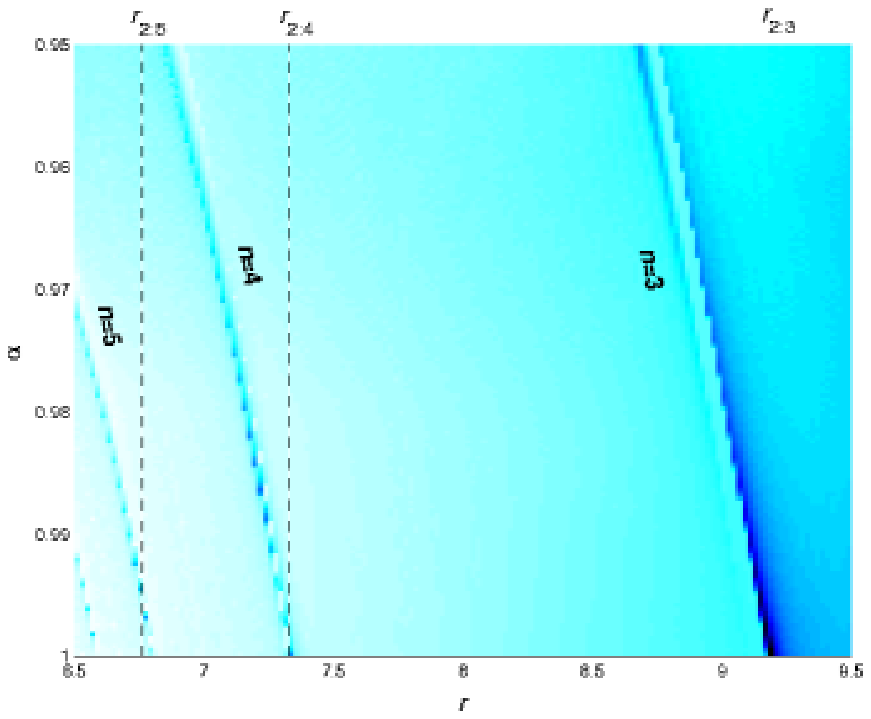}
\hfill
\FigureFile(0.48\textwidth,48\textwidth){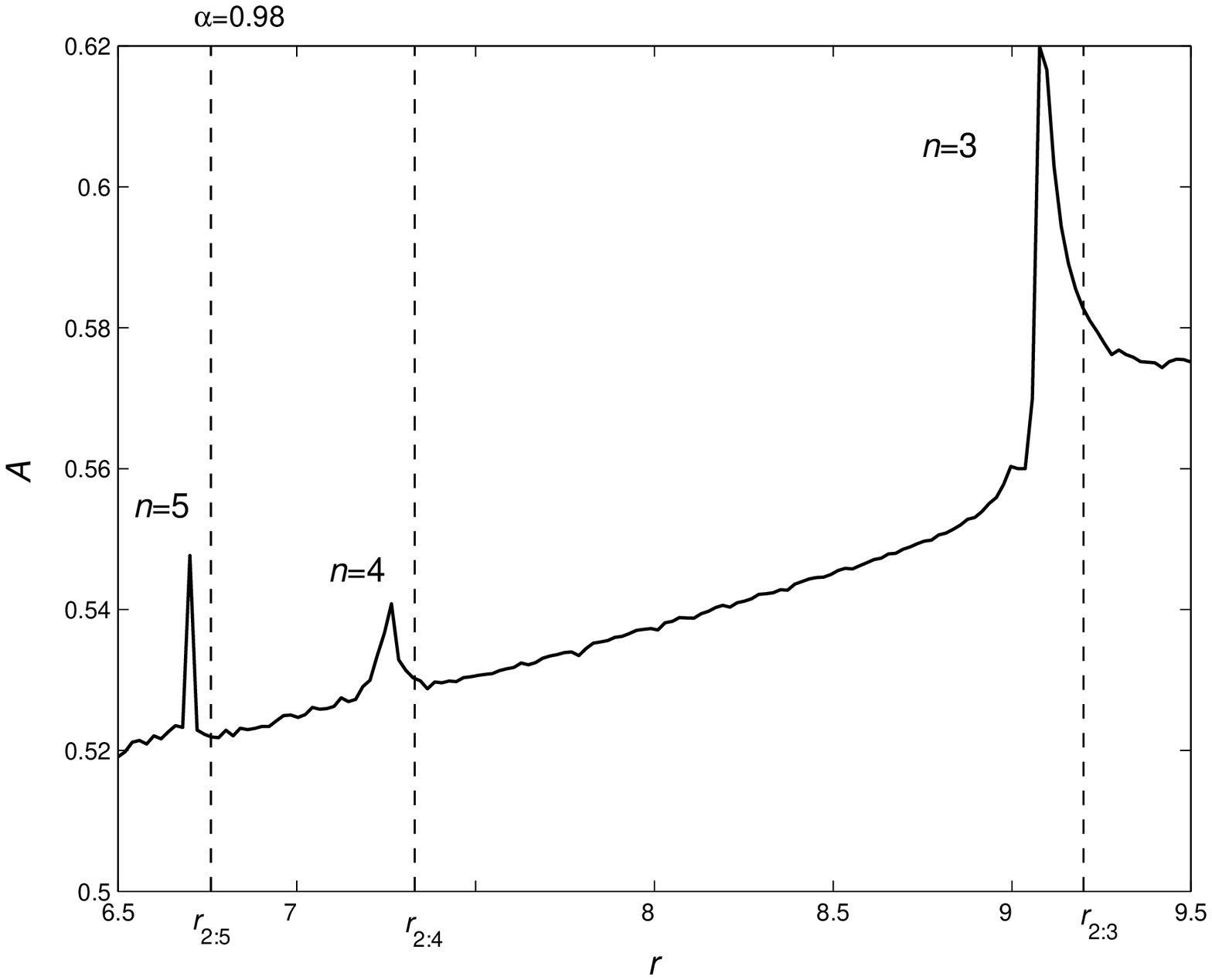}
\end{center}
\caption{Left: Location of the first three resonances is shown 
for a particular choice of initial amplitude, $\delta\theta(0)=0.5$, by
encoding the growth rate of $|\delta\theta|$ from eq.~(\ref{2.6a}) 
with different levels of shading in the $(r,\alpha)$-plane
(radius versus strength of perturbation). 
Note that $\alpha$ decreases along the $y$-axis.
For $\alpha=1.00$,  the resonant orbits are 
located at radii $r_{m:n}$ (indicated by vertical lines) where 
$\omega_r$ and $\omega_\theta$ are in rational ratios, $m:n$.
In particular, the trace of the $2:3$ resonance ($n=3$)
starts from $r_0=9.2GM/c^2$ in the Paczy\'nski-Wiita potential.  
The frequencies and radius of resonant orbits vary when $\alpha$
is changed. 
Right: The maximum amplitude $A$ of $\delta\theta(t)$ for a 
fixed value of $\alpha=0.98$ as a function of the radius of the 
unperturbed circular orbit.} 
\label{fig3}
\end{figure*}

\section{Strong gravity's $2:3$ resonance}
A parametric resonance instability occurs 
near $\omega_r = {2 \omega_{\theta}/ n}$ for $n=1,2,3,...$,
in an oscillator that obeys the Mathieu type equation of 
motion (Landau \& Lifshitz 1976),

\begin{equation}
{\ddot {\delta \theta}} + \omega_{\theta}^2 \left [ 1 
+ h_1 \cos (\omega_r t)\right ]\delta \theta + \lambda\,\delta {\dot \theta}=0.
\label{8}
\end{equation}
When the coupling is weak ($h_1$ small),
the strongest instability occurs for the lowest value of $n$ possible
(see Figure~\ref{fig2}). 

With this textbook mathematics in mind,
let us consider a simplified version of eqs.\ 
(\ref{2.6a})--(\ref{2.6c}), with
$f_i=0$, and some of the higher order terms neglected. With accuracy up to
linear terms in (\ref{2.6b}) one obtains, obviously, 
a solution $\delta r(t)\propto\cos\omega_r t$.
Substituting this solution in (\ref{2.6a}) brings this equation 
close to the standard form (\ref{8}).
We have solved these equations for $n\omega_r=2\omega_\theta$,
with $n$ a positive real parameter. 
The instability regions for a lower-order version of equations 
(\ref{2.6a})--(\ref{2.6b})  resemble the well-known tongues
of instability of the Mathieu equation.
Since $\omega_r < \omega_{\theta}$ in general relativity,
the lowest value for which the resonance can occur is $n=3$,
and the first two tongues of instability in Fig.~\ref{fig2} will be 
absent; i.e., the ratio of the two eigenfrequencies in
(\ref{8}) is $2:3$ for the strongest resonance
(Klu\'zniak \& Abramowicz 2002).

However, the behaviour of the resonance can only be properly studied
if the third-order terms in (\ref{2.6a}) and the second-order 
terms in (\ref{2.6b}) are retained.
These terms provide the non-linearity necessary to saturate the amplitude
at a finite value and the damping which affects the frequency of 
oscillations. Using the standard analytic method of
successive approximations (Landau \& Lifshitz 1976), we verified
that, as expected, no parametric resonance occurs for strictly geodesic
motion, i.e., for $f_i =0$ in eqs.\ (\ref{2.6a})--(\ref{2.6c}). 
We also checked numerically that there is
no resonance in the geodesic case.
However, the resonance does  occur (as anticipated
in Klu{\'z}niak \& Abramowicz 2002) for even very slightly
non-geodesic motion, when the higher order terms are influenced
by non-geodesic effects of a rather general form and right amplitude.

\section{Non-geodesic coupling}
In general it is not possible to
specify the form of the very small force $f_i$ exactly, because the nature of
non-geodesic forces in accretion discs is not yet very accurately known.
Here, we only explore the basic mathematical form
of the solutions, not the physical origin of
the $f_i$ force. We assume {\it ad hoc\/} that the
possible non-geodesic effects have the form of a non-linear
 isotropic coupling between the two components of geodesic deviation:
\begin{equation}
f_{\theta} = + \alpha_{\theta}\left[
{1\over 2}(\omega_{\theta}^2)^{\prime \prime} 
\delta r^2  +
{1\over {6r^2}}\left ( {{\partial^4 {\cal U}}\over {\partial
~\theta^4}}\right)_{\ell} \delta \theta^2 \right] \delta \theta,
\label{9}
\end{equation}
\begin{equation}
 f_r = - \alpha_r 
 \left[1-{{\delta r}\over  r }\right ] 
 r {\left(\dot{\delta\theta}\right)^2}.
 \label{10}
\end{equation}
Here, $\alpha_r$ and $\alpha_\theta$ measure the strength of
non-geodesic forces in comparison with the geodesic terms of corresponding
form and order. We first assumed $\alpha_r=\alpha_{\theta}\equiv\alpha$ 
for the sake of simplicity.\footnote{Perturbations considered here
may be consistent with the formation of non-axially
symmetric coherent structures in accretion discs (``planets'',
``vortices'', ``magnetic flux tubes'' or ``spiral waves''),
as proposed by several authors: Abramowicz et al.\ (1992); 
Goodman, Narayan, \& Goldreich (1987); Adams \& Watkins (1995);
Bracco et al.\ (1998); Karas (1999); Kato (2002, 2003).}

After choosing a particular value of $\alpha$, we
solved eqs.\ (\ref{2.6a})--(\ref{2.6b}) numerically by using
the Runge-Kutta adaptive step-size routine. The integration was performed 
over a wide range of starting radii, $6\le{}r_0(c^2/GM)\lesssim15$.
For the initial conditions, we took $\delta\theta(0)=0.5$,
$\delta\dot{\theta}(0)=0.01$, $\delta{}r(0)=0.5$, and $\delta\dot{r}(0)=0.01$. 
We then repeated the same procedure for another value of $\alpha$, 
and in this way we constructed a sequence of solutions that is 
parameterized by $\alpha$ and $r_0$.

One can locate radii $r_{_{\rm{}P}}(\alpha)$ for which
$|\delta\theta(t;r_{_{\rm{}P}};\alpha)|$ 
grows in time up to a maximum amplitude. The saturation
amplitude, $A$, is
shown in Fig.~\ref{fig3} (right panel) as a function of radius.
Clearly, the largest amplitudes correspond
to parametric resonance.  Close to those
radii the test-particle epicyclic frequencies
are in rational ratios $n:m$; 
$r_{_{\rm{}P}} = r_{3:2}$, $r_{4:2}$, and $r_{5:2}$. 
Note that the growth rate of
the amplitudes $|\delta\theta(t)|$, $|\delta r(t)|$ 
decreases with decreasing $\alpha$.
The $2:3$ resonance is the strongest one, as expected. 

The exact slope of the $2:3$ resonance trace 
$r_{_{\rm{}P}}(\alpha)$, starting
from the point $(r_0,\alpha)=(9.2,1)$, 
depends on the choice of coupling parameters in eqs.~(\ref{9})--(\ref{10}), 
so the above assumption of identical values for $\alpha_\theta=\alpha_r$ is not 
universal. For example, assuming $\alpha_r=\kappa\alpha_\theta$ and 
setting $\kappa=0.2$, one finds that the end of the trace is moved to
the point $(9.0,0.95)$. It is thus shifted in the $(r,\alpha)$-plane
with respect to the case $\kappa=1$, for which the trace ends at
the point $(8.8,0.95)$, as shown in the left panel of Fig.~\ref{fig3}. 
Also the initial conditions for $\delta\theta$ and $\delta{r}$
have an influence on the trace slope. Therefore, the corresponding frequencies 
that are present in the solutions for $\delta{r}(t)$, 
$\delta{\theta}(t)$ depend on the details of the solution. Nevertheless,
the ratio of the frequencies stays at $2:3$ at the maximum amplitude. 
This fact is in agreement with 
the intuition about the properties of eqs.~(\ref{2.6a})--(\ref{2.6b}), 
and we checked it by Fourier-analyzing their solution.

In this paper we assumed the gravitational potential of eq.~(\ref{6})
for the sake of simplicity. When a more accurate, full formulation in 
general relativity is developed together with a physical mechanism 
responsible for the non-geodesic coupling term
(\ref{9})--(\ref{10}), one will be able 
to constrain the absolute value of the mass of the accreting body.

\section{Frequency correlation in Sco X-1}
As noted in Section~1, the frequency ratio of the QPOs
in some black hole sources is accurately $2:3$.
Specifically, in GRO J1655-40 the frequency ratio is that of
300~Hz and 450~Hz,
while in XTE J1550-564 it is that of 184 Hz and 276 Hz
(Abramowicz \& Klu\'zniak 2001, Remillard et al.\ 2002).
The properties of parametric resonance, discussed above,
provide a natural explanation for this ratio.

\begin{figure}[!tb]
\begin{center}
\FigureFile(0.48\textwidth,0.5\textwidth){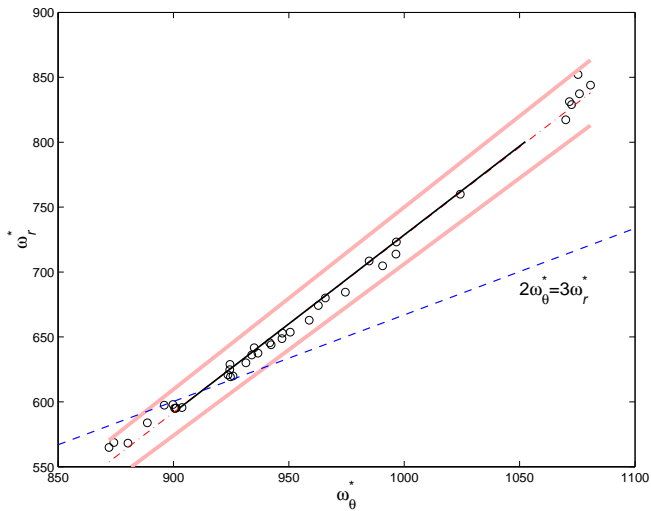}
\end{center}
\caption{Pairs of oscillation frequencies at points along the $n=3$
trace of Fig.~\ref{fig3} were obtained by integrating 
eqs.~(\ref{2.6a})--(\ref{2.6b})
with a different initial value $\delta(0)=0.8$, yielding
a frequency ratio differing from $2:3$.
The solid line is a result of the calculation 
described in the text.
For comparison with observations, we have scaled all computed
frequencies with one arbitrary multiplicative factor.
The frequencies then depend only on the strength of perturbation $\alpha$,
running in the figure from $\alpha=1.0$ at $\omega_\theta^*=900\,$ Hz
to $\alpha=0.95$ at about 1050 Hz.
Note the agreement between the slope
of the computed line and the observed kHz QPO frequencies in Sco X-1,
also shown (as circles).
The dot-dashed line is the least-squares best-fit to the data points
(two lines with a 3\% offset from the best-fit line are also plotted).
The dashed line gives a reference slope of 2/3 for comparison.
The Sco X-1 data (van der Klis et al.\ 1997)
has been kindly provided by Michiel van der Klis.}
\label{fig4}
\end{figure}

Can the same mechanism be responsible for the observed frequencies
of the kHz QPOs in neutron stars?
In neutron star sources the two kHz QPOs
vary considerably in frequency, and their ratio is not always $2:3$.
A case in point is Sco X-1, whose two kHz frequencies
are linearly correlated (Fig.~\ref{fig4}), 
but not directly proportional to each other
(van der Klis et al.\ 1997). However,
the distribution of points along the line of correlation is not uniform,
and the resulting distribution of frequency ratios has a prominent peak
at about $2:3$ value, strongly suggesting the presence of a non-linear 
resonance (Abramowicz et al.\ 2002b).

We note that QPOs are not coherent oscillations, and that the typical
integration time (minutes) greatly exceeds the coherence time of the signal
(fraction of a second).
We find that the slope of Sco X-1 frequency correlation
can be reproduced if we assume that the signal observed 
at a given time corresponds to the frequencies originating at a particular
radius, for example the radius $r_{_{\rm{}P}}(\alpha)$ discussed in the previous
section. If this is {\em{}not\/} strictly equal to the radius where the maximum 
amplitude is attained, the ratio of the frequencies departs from 2/3.
Specifically, taking an arbitrary initial amplitude $\delta\theta(0)$
we constructed Fourier power spectra of 
$\delta \theta (t; r_{_{\rm{}P}},\alpha)$,
and $\delta r (t; r_{_{\rm{}P}},\alpha)$. These power spectra peak at 
frequencies $\omega_\theta^*(\alpha)$ and $\omega_r^*(\alpha)$,
which we interpret here as the observed pair of oscillations.

Figure~\ref{fig4} shows the correlation between $\omega_\theta^*$
and $\omega_r^*$ along the $\alpha$-trace from Fig.~\ref{fig3}. 
We assumed $\kappa=1$ as before. 
The actual value of $\delta\theta(0)$ 
has been selected to match the slope of Sco X-1 data (van der
Klis 1997). Because $\alpha$ measures
the deviation from geodesicity, its value is likely to be a function of 
the accretion rate in the disc, $\alpha ({\dot M})$. This notion
agrees with the reported correlation between the frequencies and the
position of the source along the {\sf{}Z} curve in the color-color diagram.
Fig.~\ref{fig4} illustrates that the simple scheme that we introduced
in the present paper may, indeed, explain the slope of 
$\omega_r^*(\omega_\theta^*)$ relation.

\section{Conclusions}
We have expanded, through the third order, relativistic equations of 
test motion about a circular geodesic, and have shown
that the lowest order expansion corresponds to Mathieu's equation
with non-standard damping. As expected, the solutions
of these (unperturbed) equations do not show
any resonant phenomenon---they simply describe geodesic motion.

When a sufficiently large non-geodesic perturbation is imposed, 
the deviations grow rapidly at certain resonant radii
and the motion departs from circular geodesics.
This is caused by parametric resonance between the meridional
and radial epicyclic motions, and the strongest resonance occurs
when the two dominant frequencies are 
near ratio $2:3$. 

\medskip
We thank Michiel van der Klis for providing the Sco X-1 data and for a 
very helpful discussion. Most of this work was done during MAA's and 
VK's visit to the supercomputer centre UKAFF at Leicester University, 
and MAA's, WK's and WHL's visits to IAP (Paris) and SISSA (Trieste).
WK held a post rouge at CESR, funded by CNRS. VK acknowledges support 
from GACR 205/03/0902 and GAUK 188/2001, and WHL from CONACYT (36632E). 
The Center for Particle Physics is supported by the Czech Ministry of 
Education Project LN00A006.

\end{document}